\documentclass[runningheads]{llncs}
\usepackage{hyperref}
\usepackage{makecell}
\usepackage{comment}
\usepackage{booktabs}
\usepackage{multirow}
\usepackage[disable]{todonotes}
\usepackage{tikz}
\usetikzlibrary{arrows.meta}  
\usetikzlibrary{positioning}   
\usetikzlibrary{backgrounds, positioning, shapes, calc} 
\usepackage{xcolor}
\usepackage[T1]{fontenc}

\usepackage{algorithm}
\usepackage{algpseudocode}
\usepackage{listings}
\usepackage{graphicx}

\usepackage{amsmath}

\begin{document}
\title{Inferring Sensitive Attributes from Knowledge Graph Embeddings: Attack and Defense Strategies}
\titlerunning{Attribute Inference Attack on KGE}

\author{Yasmine Hayder}
\authorrunning{Y Hayder et al.}

\institute{LIFO, INSA CVL, Univ. Orléans, Inria, France\\
\email{yasmine.hayder@insa-cvl.fr}\\
}
\maketitle      
\begin{abstract}
Knowledge Graphs (KGs) are a powerful representation of linked data, offering flexibility, semantic richness, and support for knowledge enrichment and reasoning. They help data owners organize and exploit heterogeneous data to provide insightful services (e.g., recommendations), yet real-world KGs are often incomplete, hiding true facts or missing valuable insights. Knowledge graph embedding techniques are commonly used to infer valuable missing information. However, reasoning over KGs can inadvertently expose sensitive user information, even when such data is not explicitly stored. In this work, we investigate the privacy risks associated with KGE-based reasoning, focusing on attribute inference attacks where adversaries attempt to deduce sensitive user attributes from seemingly non-sensitive outputs. We propose and evaluate a framework that mitigates these privacy risks by applying post processing sanitization techniques to KGE outputs.  Preliminary results demonstrate the effectiveness of these attacks on the outputs of KGE models, and explore the trade-off between recommendation quality and privacy protection when applying randomization based approaches, highlighting the need to experiment with more advanced techniques in future work to address this issue.

\end{abstract}

\section{Introduction}

The Semantic Web aims to transform the web into a machine-interpretable space by enabling data to be shared, linked, and reused across applications and domains. At the core of this vision lie Knowledge Graphs (KGs)~\cite{fensel2020introduction}, which provide a structured and semantically rich representation of entities and their relationships using formal vocabularies. By organizing heterogeneous data in a unified framework, knowledge graphs support advanced reasoning capabilities, allowing systems to infer implicit knowledge from explicitly stated facts. Knowledge Graph Embedding (KGE) models~\cite{ge2024knowledge} play a crucial role in addressing the inherent incompleteness of knowledge graphs, enabling reasoning over large, sparse datasets. They have been successfully applied in domains such as search engines~\cite{mai2018combining}, recommender systems~\cite{zhang2024review}, and biomedical research~\cite{chang2020benchmark}. However, this growing reliance on reasoning over linked data introduces significant privacy concerns~\cite{chen2024survey}. Unlike traditional relational databases, KGs encode rich semantic dependencies between entities and their attributes, which can be exploited to infer sensitive information even when direct access to such data is restricted. In link prediction, there is an incentive for the data holder to fill some of those missing links for efficiency reasons, compounding the issue. Consider a practical example in the context of an e-commerce platform. A practitioner uses a knowledge graph constructed from user–item interactions to recommend products to customers. Some users have consented to their personal data being used to improve recommendations, while others have opted out of sharing sensitive information such as age or gender. The challenge arises when the system generates recommendations based on user histories: even without explicit access to sensitive attributes, the pattern of recommended items or past interactions could allow the practitioner to infer private information about users who did not consent. For instance, if a user consistently interacts with products typically favored by a particular demographic, the system might inadvertently reveal their sensitive attribute, violating their privacy. The main goal of this research is  to enable effective reasoning while ensuring that sensitive attributes remain protected.
\todo{this is doubly true when this attribute is used by the KGE for the recommendation}

Addressing privacy in knowledge graph reasoning is a major challenge~\cite{chen2024survey}. Existing work mainly focuses on protecting against white box attackers (having full access to the model’s code, parameters, and embeddings) by adding privacy mechanisms at the training stage \cite{han2022framework,chen2021fede,zhang2021graph}. However, this assumption is often too strong for real-world applications, where attackers typically operate in a black-box setting (can only observe the model’s outputs) and interact with models through APIs.
In this paper, we present preliminary work to see if it is, in this setting, more efficient to preserve full model utility and introduce a privacy layer at the deployment stage. Unlike training time sanitization methods, which are often tailored to specific model architectures, our approach is model agnostic and can be applied to any system that outputs a ranked list.

\section{Background and Related Works}
In this section, we provide the background on knowledge graph embeddings and attribute inference attacks necessary to the foundations of this paper. Then we proceed by related works and positioning of our contributions compared to it.

\subsection{Background}
\todo{if it matters, are KGE usually on a small variety of relations?}

\textbf{Knowledge graphs (KG)} represent knowledge as a set of triples 
$(h,r,t)$, where $h$ and $t$ denote entities and $r$ denotes the relation between them. While this representation is expressive, real-world KGs are often incomplete. To address this, recent work has shifted toward knowledge graph embeddings (KGE).

\noindent\textbf{Knowledge graph embeddings (KGE)} aim to represent entities and relations of a knowledge graph as vectors in a continuous low-dimensional space, allowing them to model heterogeneous and diverse sets of entities and relations. The goal is to learn representations in which valid triples receive high plausibility scores while invalid ones receive low scores. By transforming symbolic graph data into numerical vectors, KGE models enable reasoning tasks such as link prediction and query answering to be performed through efficient numerical computation. During training, embeddings are typically initialized randomly and optimized using a scoring function that measures how well a triple $(h,r,t)$ is represented in the embedding space. 

One of the earliest and most influential KGE models is TransE~\cite{bordes2013translating}, which represents relations as translations in the embedding space. For a triple $(h,r,t)$, TransE models the relation $r$ as a vector translation from the head entity $h$ to the tail entity $t$, enforcing the relation $h + r \approx t$. Despite its simplicity, TransE is effective at modeling one-to-one relations but struggles with more complex patterns such as one-to-many, many-to-one, and many-to-many relationships.
Among later developments, RotatE~\cite{sun2019rotate} is a representative distance-based model that interprets relations as rotations in a complex vector space. For a triple $(h,r,t)$, the model learns embeddings such that the tail entity is close to a rotated version of the head entity, i.e., $t \approx h \circ r$, where $\circ$ denotes element-wise complex multiplication. This geometric interpretation allows RotatE to capture important relational patterns such as symmetry, antisymmetry, inversion, and composition. At inference time, queries of the form $(h,r,?)$ are answered by ranking candidate entities according to their scores, with higher-ranked entities considered more plausible. This ranking mechanism makes KGE models particularly suitable for recommendation and knowledge completion tasks.

\noindent\textbf{Attribute inference attacks (AIA)} \cite{gong2018attribute} aim to infer sensitive attributes of entities that are not explicitly stored in the data but can be deduced from observable information. In the context of knowledge graph embeddings (KGE), preventing direct inference on sensitive queries (e.g., $(\texttt{user1}, \texttt{hasGender}, ?)$) is not sufficient. Even when such queries are blocked, so-called \emph{non-sensitive} predictions, such as recommendation queries (e.g., $(\texttt{user1}, \texttt{mightLike}, ?)$), may still leak private information. A malicious user can exploit these outputs by combining them with external or publicly available knowledge to indirectly infer sensitive attributes. For example, if all recommended movies belong to the action genre, one may intuitively deduce that the user is likely male, illustrating how sensitive information can be inferred through indirect reasoning.

\subsection{Related work}
\todo{reference netflix prize?}
Several studies have shown that knowledge graph embedding (KGE) models are vulnerable to privacy attacks, particularly membership inference attacks that exploit model behavior or learned embeddings~\cite{wang2021membership}. To mitigate these risks, a range of privacy-preserving approaches has been proposed, primarily focusing on the training phase. In federated and distributed settings, where gradients or embeddings are exchanged during learning, these methods aim to prevent leakage of sensitive information by incorporating mechanisms such as differential privacy or adversarial perturbations~\cite{hu2023quantifying}.

Beyond KGE, similar privacy concerns have been identified in graph embedding and recommendation systems. For example, the de-anonymization of users in the Netflix Prize dataset demonstrated how seemingly benign outputs can leak private information~\cite{narayanan2006break}. More recent frameworks, such as GERAI, introduce sanitization strategies to reduce information leakage while preserving model utility~\cite{zhang2021graph}.

While these approaches differ in scope and technique, they share a common focus on defending against powerful adversaries, often under white-box assumptions where the attacker has access to model parameters, embeddings, or gradients. However, such assumptions are less representative of some real-world deployments, where models are often accessed through query interfaces, making black-box attacks more practical. Importantly, the existing body of work largely concentrates on mitigating privacy risks during training, leaving inference-time vulnerabilities comparatively underexplored. This gap motivates the need for defenses that operate at inference time, sanitizing model outputs to protect against black-box attacks while still enabling high-quality predictions for legitimate users—a line of research that this work aims to tackle.

\section{Problem Statement and Contributions} \label{sec:problem}

This work studies the trade-off between recommendation utility and privacy in knowledge graph embedding (KGE) based reasoning. In particular, we focus on attribute inference attacks (AIA), where a malicious adversary infers sensitive user attributes from recommendation outputs combined with external knowledge. Our objective is to design defense mechanisms that reduce the success of such attacks while preserving the usefulness of KGE based recommendations.

\paragraph{Attribute inference attack}
We model an adversary’s ability to infer sensitive user attributes from a list of recommended items, possibly leveraging external or public knowledge. Let $\mathcal{P}$ denote external datasets, $\mathcal{U}$ the set of users, $\mathcal{A}_s$ the set of sensitive attributes, and $\mathcal{L}_K$ the list of recommendation lists. We define the attribute inference function:
\[
\mathcal{I} : \mathcal{L}_K \times \mathcal{P} \times \mathcal{U} \times \mathcal{A}_s \rightarrow \{0,1\},
\]
where $\mathcal{I}(\ell, \mathcal{P}, u_i, a_s) = 1$ if the adversary successfully infers the sensitive attribute $a_s \in \mathcal{A}_s$ of user $u_i \in \mathcal{U}$ from the recommendation list $\ell \in \mathcal{L}_K$, and $0$ otherwise. For a set of users $U \subseteq \mathcal{U}$, the attack success rate is defined as:
\[
\mathcal{I}_u = \frac{\sum_{u_i \in U} \mathcal{I}(\ell, \mathcal{P}, u_i, a_s)}{|U|},
\]
i.e., the fraction of users whose sensitive attribute is correctly inferred by the adversary.

\todo{Is the goal of this paper to present and validate the approach? If so both model sanatization and recommendation sanatization should be presented and compared as strategies, even if we only have results for one}

\paragraph{Recommendation sanitization}
To mitigate privacy leakage from attribute inference attacks, we introduce a sanitization algorithm that modifies the original recommendation list produced by a KGE model before it is released to practitioners. Formally, we define a sanitization function:
\[
\mathcal{S} : \mathcal{L}_K \times \mathcal{M} \rightarrow \mathcal{L}_K,
\]
which takes as input a recommendation list $\ell \in \mathcal{L}_K$ and the universe of items $\mathcal{M}$, and outputs a sanitized list $\ell' = \mathcal{S}(\ell, \mathcal{M}) \in \mathcal{L}_K$. The role of $\mathcal{M}$ is to provide candidate items that can be used to replace or perturb elements of $\ell$. 

\paragraph{Recommendation evaluation}
We formalize the utility of a recommendation list by comparing a sanitized list to the original output of the KGE model. Let $\mathcal{M}$ be the set of movies and $\mathcal{L}_K$ the set of lists of size $K$ over $\mathcal{M}$. We define the evaluation function:
\[
\mathcal{Q} : \mathcal{L}_K \times \mathcal{L}_K \rightarrow [0,1],
\]
where $\mathcal{Q}(\ell,\ell')$ measures how well the sanitized list $\ell'$ preserves the utility of the original list $\ell$. Values close to $1$ indicate high utility, while values close to $0$ indicate significant degradation. The definition of $\mathcal{Q}$ is application dependent and may capture criteria such as top-$K$ overlap or diversity constraints. For a set of users $U \subseteq \mathcal{U}$, the overall utility is defined as:
\[
\mathcal{Q}_u = \frac{1}{|U|} \sum_{u_i \in U} \mathcal{Q}(\ell_i, \ell'_i).
\]

\paragraph{Contribution}
Our objective is to generate a sanitized recommendation list using $\mathcal{S}$ that maximizes utility while minimizing the success of attribute inference attacks. We formulate this as a trade-off problem between recommendation quality and privacy leakage, characterized by the utility measure $\mathcal{Q}_u$ and the attribute inference accuracy $\mathcal{I}_u$. Our goal is to analyze how different sanitization strategies balance these two quantities and to identify configurations that achieve a favorable trade-off.

\section{Research Methodology and Approach} 
This research is motivated by preliminary experiments demonstrating that a simple attribute inference attack (AIA) based on knowledge graph embeddings can infer sensitive attributes from recommendation outputs with success rates significantly higher than random guessing. Building on this observation, we first formalized the problem by defining models for attribute inference risk, recommendation utility, and output sanitization, as described in Section~\ref{sec:problem}. We then implemented and evaluated a KGE-based attack model and proposed an initial post-processing perturbation algorithm that sanitizes recommendation lists by partially replacing model outputs with items uniformly sampled illustrated in algorithm \ref{alg:sanitization} with an option to further perturb the list by shuffling the output list of the model before selecting top items. This baseline defense, illustrated in the following section, serves as a first step toward mitigating inference risks at prediction time. 
Ongoing and future work focuses on extending this study in several directions. We plan to conduct a broader empirical evaluation of diverse AIA strategies across multiple KGE models and to design more advanced sanitization mechanisms, including approaches with formal privacy guarantees based on differential privacy~\cite{dwork2006differential}\todo{potentially move this up in the method presentation? Is this just for recommendation sanatization, just for model sanatization, or both? (in which case it's ok here)}.

\begin{algorithm}
\caption{Recommendation List Sanitization $\mathcal{S}$ and $\mathcal{S}_{\text{shuf}}$}
\label{alg:sanitization}
\begin{algorithmic}[1]
\Procedure{Sanitize}{$\ell, \mathcal{M}$} 
    \Comment{$\ell$: original recommended list, $\mathcal{M}$: movie universe}
    \State $\ell_{\text{top}} \gets$ shuffle($\ell$,K) \Comment{optional shuffle 10 top predicted items for $\mathcal{S}_{\text{shuf}}$}
    \State $\ell' \gets \ell_{\text{top}}[1:t]$ \Comment{take top t items from shuffled/original list}
    \State $R \gets$ RandomSample($\mathcal{M} \setminus \ell_{\text{top}}, r$) \Comment{take random items from $\mathcal{M} \setminus \ell_{\text{top}}$}
    \State $\ell' \gets \ell' \cdot R$
    \State \textbf{return} $\ell'$
\EndProcedure
\end{algorithmic}
\end{algorithm}

\todo{shuffle is a list operator, $\cup$ is a set operator. I assume we're on lists for now. Line 5 is then a concatenation (I assume) in which case the symbol is usually $\ell' \cdot R$ afaik}

\section{Evaluation:} 
%\textcolor{red}{Describe your evaluation or evaluation plan, which is the way you (intend to) validate your hypothesis, your results, and the value of your approach.}
\textbf{Goal :} We evaluate our approach through experiments designed to quantify the trade off between recommendation utility and privacy leakage under attribute inference attacks. The objective of this evaluation is to measure how sanitizing recommendation lists affects the adversary’s ability to infer sensitive user attributes while preserving the usefulness of recommendations produced by a knowledge graph embedding based model.

\noindent\textbf{Datasets:} 
Our experiments are conducted on two benchmark datasets: the Yahoo Movies dataset, which contains approximately 200k triples and about 7k users, and the MovieLens 100K dataset, which includes 100k triples and about 1k users. Both datasets are transformed into a standardized knowledge graph format for recommender system research using RecKG \cite{kwon2024reckg}. 
The datasets include user–movie interactions as well as demographic information such as gender, birth year, and occupation. These demographic attributes are treated as sensitive information that an adversary attempts to infer from recommendation outputs. A knowledge graph embedding (KGE) model is trained on each dataset to generate top-$K$ recommendation lists for each user, which serve as the baseline (non-sanitized) recommendations.

\noindent\textbf{Privacy leakage : } is evaluated through an attribute inference attack based on a KGE model. The attacker exploits correlations between recommended movies and demographic attributes encoded in the knowledge graph to infer sensitive information about users. The attack model observes only the list of recommended movies associated with each user and predicts the most likely value of the sensitive attribute. An attack is considered successful when the correct attribute value is ranked ahead of incorrect alternatives, meaning that the attacker’s top prediction matches the true user attribute. The overall inference success rate is computed as the fraction of users for which this condition holds. Attribute inference is classically studied using supervised classification models such as k nearest neighbors or other learning based predictors. In this work we focus exclusively on a KGE based attack for their simplicity. A systematic comparison between KGE based attribute inference and classification based attacks is left for future work.

\noindent\textbf{Sanitization : } We apply our sanitization algorithm to the recommendation lists produced by the KGE model by replacing a controlled portion of recommended movies with randomly sampled items from the movie universe as illustrated in Algorithm~\ref{alg:sanitization} with an option to shuffle the recommended list produced by the KGE to perturb items' ranks\todo{with/without shuffle or just with?}. By varying this proportion, we generate recommendation lists with different privacy and utility characteristics.

\noindent\textbf{Utility measure: }We define recommendation utility as $\mathcal{Q}_u$, the average across users of the ratio between the sum of model assigned scores for the sanitized list and that of the top-$K$ items from the original KGE model. This score based measure captures how much of the model’s original recommendation quality is retained after sanitization.

\noindent\textbf{Future Evaluation : }The current evaluation assumes an attacker that observes only a recommendation lists \todo{In that case the final recommendation of the algorithm would be a set and not a list, but its name is $\ell'$. If there is a cast, it should be made explicit in comments and notations} and has no access to users’ historical interaction data. Future work will consider more powerful adversaries that can exploit interaction histories in addition to the suggested recommended items produced by a KGE. 

\section{Results:} 
We use a recommendation model trained with the RotatE knowledge graph embedding model. This choice is motivated by the characteristics of our dataset: it contains both one-to-many relations (e.g., users to movies) and one-to-one relations (e.g., users to demographic attributes). RotatE is well-suited for modeling such diverse relational patterns. While our framework could be applied to other KGE models, experiments showed that RotatE provided strong recommendation performance in our experiments.
 The model achieved a Top-5 recommendation accuracy of 0.22 and a Top-10 accuracy of 0.30.
For the attribute inference attack, we used a separate RotatE model with smaller capacity (embedding dimension 70, 100 epochs, and batch size 64), as this smaller model is sufficient to capture sensitive patterns even with low resource use. An attack was considered successful if the correct value of a sensitive attribute (e.g., gender or age group) was ranked higher than false candidates.
For the gender attribute, users were classified as male or female, making a random guess baseline around 0.5. To test the attack models, we selected 10\% of users from the dataset and retrieved their sensitive attributes from the training set and place them in the test set to evaluate the model’s ability to infer sensitive information from non-sensitive attributes. 

\todo{Display preference (up is better/down is better) and standard deviation in the table or its caption. Put results that represent a significative best/improvement in bold.}

Table~\ref{tab:1} reports results showing the trade-off between inference success and recommendation utility for the \textit{gender} attribute. Experiments for the other sensitive attributes are left for future works. The table presents the attribute inference accuracy $\mathcal{I}_u$ under different levels of sanitization, alongside the corresponding recommendation quality $\mathcal{Q}_u$, as we vary the number of retained top recommendations $t$ and the number of randomly selected items $r$.

\textbf{Interpretation: }For the gender attribute, we observe that the attack success decreases as more random items are introduced, at the cost of reduced recommendation quality. This suggests that even a simple sanitization strategy can effectively influence the privacy–utility trade-off. However, measuring utility solely through $\mathcal{Q}_u$ may underestimate the usefulness of randomly added items, as some may still align with the user’s broader interests beyond the sensitive attribute. Furthermore, results with shuffled recommendation lists $\mathcal{S}_{\text{shuf}}$ indicate that item ordering also plays a role, as shuffling slightly reduces inference success. Overall, these findings suggest that while the current approach is effective, it remains coarse, and motivate the exploration of more refined strategies either by selecting replacement items in a more informed way (e.g., popular or trending items) or by leveraging mechanisms such as the exponential mechanism \cite{mcsherry2007mechanism} to better balance privacy preservation and recommendation utility.

We provide herein \href{https://github.com/Yasmine-Hayder/Attribute-inference-attack-Knoweldge-graph-Embedding/}{https://github.com/Yasmine-Hayder/Attribute-inference-attack-Knoweldge-graph-Embedding/}
 a code implementation of our algorithms and the data used.
\begin{table}
\centering
\caption{Privacy–utility trade-off under sanitization with and without shuffle. $\mathcal{I}_u$ denotes inference accuracy and $\mathcal{Q}_u$ denotes recommendation quality. $t$ is the number of top movies recommended by the model, while $r$ is the number of randomly selected movies drawn from the universe $\mathcal{M}$. The experiments were repeated five times, each with a randomly selected subset of users, and the table shows the mean results across these runs with standard deviation between [0.01,0.02] for Yahoo Movies Dataset and [0.03,0.04] for MovieLens Dataset.}
\small
\label{tab:1}

\setlength{\tabcolsep}{5pt}
\renewcommand{\arraystretch}{1.25}

\begin{tabular}{c ccccc | ccccc}
\hline
\multirow{2}{*}{Metric}
& \multicolumn{5}{c|}{5 Recommendations}
& \multicolumn{5}{c}{10 Recommendations} \\
\cline{2-6} \cline{7-11}

& Top  & t3–r2 & t2–r3 & t1–r4 & Rand
& Top  & t7–r3 & t5–r5 & t3–r7 & Rand\\
\hline

\multicolumn{11}{c}{\textbf{Yahoo Movies Dataset}} \\
\hline

$\mathcal{I}_u\;(\mathcal{S})$
& 0.70  & 0.67 & 0.65 & 0.61 &  \multirow{2}{*}{ 0.52}
& 0.71  & 0.69 & 0.68 & 0.66 &  \multirow{2}{*}{ 0.51} \\

$\mathcal{I}_u\;(\mathcal{S}_{\text{shuf}})$
& 0.68 &  0.64 & 0.60 & 0.60  & 
& 0.70 &  0.67 & 0.67 & 0.64  &  \\

\hline 

$\mathcal{Q}_u\;(\mathcal{S})$
& 1  & 0.74 & 0.61 & 0.47 & \multirow{2}{*}{0.32}
& 1  & 0.81 & 0.68 & 0.55 & \multirow{2}{*}{ 0.34} \\

$\mathcal{Q}_u\;(\mathcal{S}_{\text{shuf}})$
& 0.96  & 0.70 & 0.58 & 0.45 & 
& 1 & 0.80 & 0.67 & 0.53 & \\

\hline

\multicolumn{11}{c}{\textbf{MovieLens Dataset}} \\
\hline

$\mathcal{I}_u\;(\mathcal{S})$
& 0.63 & 0.61 & 0.61 & 0.57 & \multirow{2}{*}{0.53}
& 0.71 & 0.65 & 0.60 & 0.57 & \multirow{2}{*}{0.52} \\

$\mathcal{I}_u\;(\mathcal{S}_{\text{shuf}})$
& 0.64 & 0.59  & 0.59 & 0.57  & 
& 0.65 & 0.61 & 0.61 & 0.59 &  \\

\hline

$\mathcal{Q}_u\;(\mathcal{S})$
&  1 & 0.81 & 0.71 & 0.62 & \multirow{2}{*}{0.51}
&  1 & 0.86 & 0.77 & 0.67 & \multirow{2}{*}{0.52} \\

$\mathcal{Q}_u\;(\mathcal{S}_{\text{shuf}})$
& 0.98 & 0.79 & 0.69 & 0.61 & 
& 1 & 0.85 & 0.76 & 0.67 & \\

\hline
\end{tabular}
\end{table}
\newpage
\section{Conclusions}
This work highlights the privacy risks associated with knowledge graph embedding (KGE) models, showing that sensitive attributes such as gender can be inferred from recommendation outputs with accuracy well above random guessing in a black-box setting. We study the efficiency of a post-processing sanitization strategy that perturbs recommendation lists by combining top-ranked items with randomized ones to mitigate such risks. Our findings answer our primary research question: whether sensitive attributes can be inferred from KGE-based recommendations and mitigated at inference time. The results confirm both the existence of this risk and the effectiveness of output perturbation as a practical defense, without requiring changes to the training process.

\textbf{Future works : }
One promising line of work involves augmenting knowledge graph embedding models with an additional layer of inference rules, either mined \cite{galarraga2013amie} automatically from the dataset or explicitly provided by data owners. Recent approaches, such as RulE~\cite{tang2024rule} and RNNLogic~\cite{qu2020rnnlogic} improve predictive performance through logical rules, they may also increase privacy leakage by strengthening correlations with sensitive attributes. We therefore aim to investigate how such reasoning layers impact attribute inference attacks, and to explore the use of Onto-DP \cite{hayder2026onto} in this context, as it is specifically designed to provide differential privacy guarantees in ontology-based reasoning systems.

\subsubsection{\ackname}Special thanks to my supervisors Adrien Boiret, Cédric Eichler
and Benjamin Nguyen for their help and guidance.

\bibliographystyle{unsrt}
\bibliography{sample-base}

\end{document}